\documentclass[aip,apl,twocolumn,preprintnumbers,amsmath,amssymb]{revtex4}

\usepackage{graphicx}
\usepackage[colorlinks=true,citecolor=blue]{hyperref}

\begin{document}
    \title{Design and analysis of photonic crystal coupled cavity arrays for quantum simulation}
    \author{Arka Majumdar}
    \email{arkam@stanford.edu}
    \author{Armand Rundquist}
    \author{Michal Bajcsy}
    \author{Vaishno D. Dasika$^\dag$}
    \author {Seth R. Bank$^\dag$}
    \author{Jelena Vu\v{c}kovi\'{c}}
    \affiliation{E.L.Ginzton Laboratory, Stanford University, Stanford, California, $94305$\\
    $^\dag$ Microelectronics Research Center, University of Texas, Austin, Texas, $78712$}
\begin{abstract}
We performed an experimental study of coupled optical cavity arrays in a photonic crystal platform. We find that the coupling between the cavities is significantly larger than the fabrication-induced disorder in the cavity frequencies. Satisfying this condition is necessary for using such cavity arrays to generate strongly correlated photons, which has potential application to the quantum simulation of many-body systems.
\end{abstract}
\maketitle
\section{Introduction}
Solving strongly correlated quantum many-body systems is a formidable task. One promising approach is to mimic such complicated systems using another simpler and easily controllable quantum system, as envisioned by Richard Feynman \cite {feynman_article}. To that end, the first demonstration of quantum phase transitions with ultra-cold atoms in an optical lattice \cite{bloch_qpt} sparked a significant amount of  research on quantum simulation with atomic systems \cite{bloch_review}. Another very promising direction of using photons themselves as the interacting particles has generated considerable interest recently \cite{ciuti_fluid_light}. The main idea of this approach is to obtain a correlated ``quantum fluid of light'' \cite{ciuti_fluid_light} by building a coupled network of nonlinear electromagnetic cavities. The photons can hop between cavities due to the electromagnetic coupling and can repel each other in the same cavity due to the intra-cavity nonlinearity. We note that such a coupled cavity network exhibits rich physics such as topologically protected optical delays \cite{hafezi_topology}, even without any nonlinearity, although having nonlinear cavities opens up many more avenues of research. Obviously, the optical nonlinearity required for significant repulsion at low photon number is very high, and in current technology, only $2$-level systems (for example, atoms, single quantum emitters such as quantum dots (QDs) or super-conducting transmon qubits) strongly coupled to a cavity provide such strong nonlinearity in the photon blockade regime \cite{arka_tunneling, blockade_imamog,birnbaum_nature,AF_natphys}. In most of the applications relating to quantum simulations, one needs to deterministically position single quantum emitters in each of the cavities, which is very difficult to achieve in the state-of-the-art solid-state technology. However, recently several groups have demonstrated deterministic positioning of self-assembled QDs \cite{kapon_NRDC}, and the hope is that these site-controlled QDs will also perform well within the setting of cavity quantum electrodynamics (CQED). Another approach would be to use a bulk nonlinearity or quantum well nonlinearity, but significantly enhanced by a cavity with high quality (Q) factor and low mode volume \cite{gerace_bulk, carusotto_imamoglu}. We note that such a platform consisting of coupled nonlinear cavities is useful not just for the quantum simulation, but also for quantum error correction \cite{joe_qecc} as well as for classical optical signal processing \cite{mabuchi_opt_logic}.

Although plenty of theoretical proposals for simulating interesting physics in such a coupled cavity array (CCA) are present in the literature, the experimental progress in that direction is rather limited. As one needs to have many cavities for this operation, a solid-state system is obviously an ideal choice. However, due to imperfect nano-fabrication solid-state cavities have inherent disorder, resulting in different resonance frequencies than the cavities were originally designed for. Such disorder might limit the utility of CCAs for quantum simulation. However, in a recent paper it is argued that as long as the coupling strengths are much larger than the disorder, the CCAs can be used for quantum simulation, and it is shown that microwave transmission line cavities for circuit QED satisfy this condition \cite{houck_cqed_paper}.

In this paper, we demonstrate high-Q $2$-D CCAs based on photonic crystals fabricated in GaAs with embedded high denisty self-assembled InAs QDs. Although a pair of coupled cavities, also known as a photonic molecule, is well studied in the literature \cite{phot_mol_AM,kapon_2011, Intoni_2011}, relatively little literature exists for CCAs. A $2$-D CCA of photonic crystals in GaAs (with multiple quantum wells as active materials) has been studied previously for increasing the output light intensity from nano-lasers or slowing down light \cite{hatice_cca_opex,hatice_1,hatice_2,EL_denner}, but the Q-factors of the cavities were too low to identify individual cavity modes. A long chain of high-Q coupled cavities has been studied in silicon \cite{notomi_array_cavity}, but the physical phenomena observable in such a $1$-D chain are rather limited. While a $1$-D array \cite{book_quant_1D} has been studied as a platform to simulate the physics of Bose glass \cite{bose_glass_1D}, and Tonks-Girardeau gas \cite{chang_1D}, a $2$-D array is a more suitable candidate for simulating many other systems including topologically non-trivial states such as the fractional quantum Hall state \cite{FQHE_greentree,FQHE_bose,FQHE_carusotto}.

\section{Spectra of coupled cavities}
\begin{figure}
\includegraphics[width=3.5in]{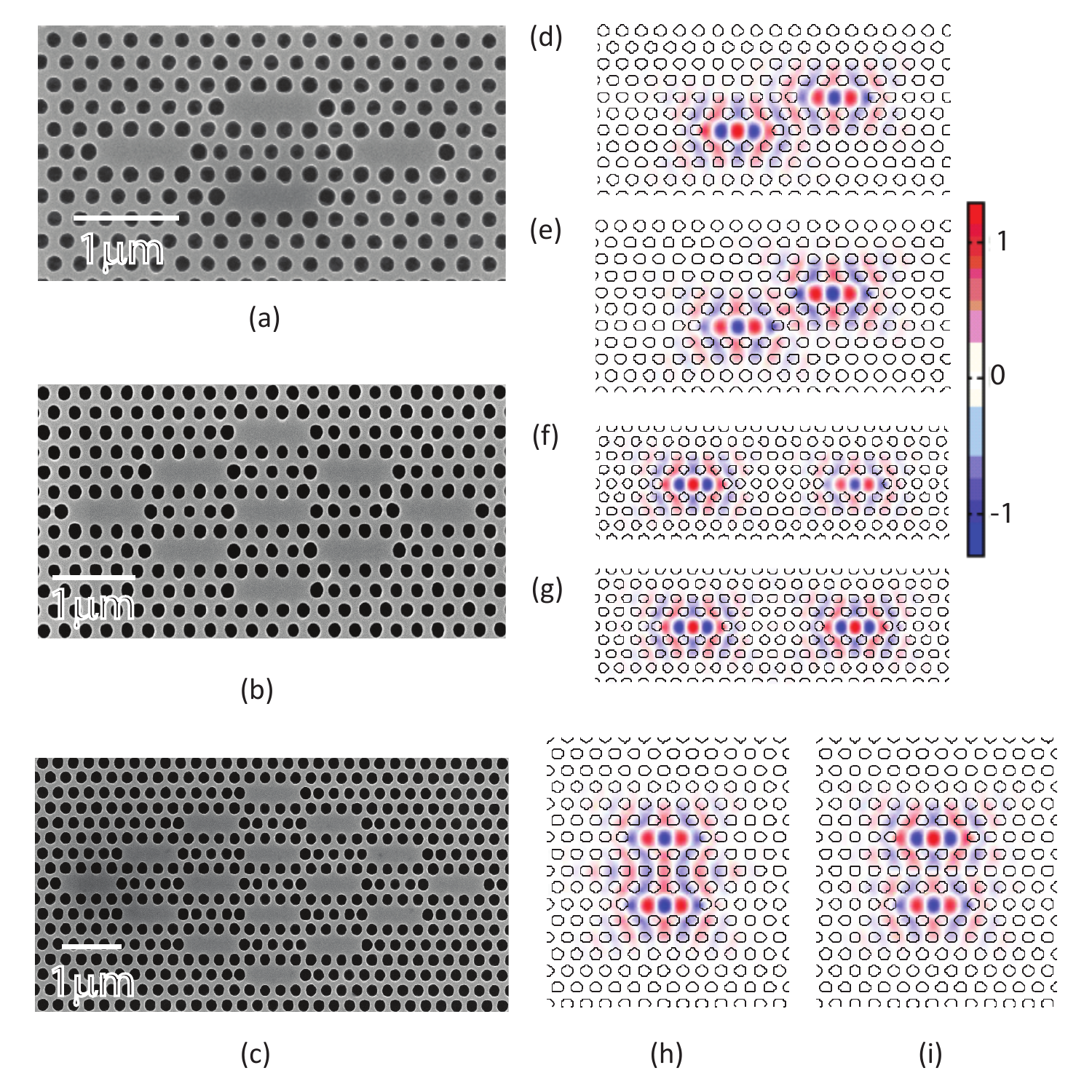}
\caption{(color online) Scanning electron micrograph (SEM) images of CCA with (a) $4$ cavities, (b) $9$ cavities and (c) $16$ cavities. The simulated electric field profiles for each of the two super-modes of the coupled cavities are shown: (d),(e) for $60^o$ coupled cavities; (f),(g) for laterally coupled cavities; (h),(i) for vertically coupled cavities.
\label{Fig1_SEM_multi_cavity}}
\end{figure}

In our experiment, we employ an array of linear three-hole ($L3$) defect photonic crystal cavities, typically studied in single QD-cavity QED experiments \cite{article:eng07,article:hen07}. The fundamental mode of such a cavity is linearly polarized in the direction orthogonal to cavity axis; in our proposed CCA geometry all the cavities have parallel axes, and their modes have the same polarization. The photonic crystal CCA is fabricated in a $164$ nm thick GaAs membrane (with self-assembled InAs QDs embedded at a depth of $82$ nm from the surface) using electron-beam lithography and reactive ion etching \cite{article:eng07}. Scanning electron micrographs of the fabricated structures are shown in Figs. \ref{Fig1_SEM_multi_cavity}a,b,c. Three different CCAs are designed consisting of $4$, $9$ and $16$ cavities. These cavities are coupled to each other by three different coupling strengths depending on the relative orientation and separation of two cavities. When two cavities are coupled at an angle of $60^o$ (Figs. \ref{Fig1_SEM_multi_cavity}d,e) the coupling strength $t$ is strongest; for vertically stacked coupled cavities (Figs. \ref{Fig1_SEM_multi_cavity}h,i) the coupling strength $J_1$ is smaller than $t$; and for horizontal coupled cavities (Figs. \ref{Fig1_SEM_multi_cavity}f,g) the coupling strength $J_2$ is much smaller than $t$ and $J_1$ (the difference in coupling strengths is a result of the different radiation patterns of the cavity modes, and their different overlaps in various  directions). From the finite difference time domain (FDTD) simulations we can calculate the field profiles of the coupled cavities (Figs. \ref{Fig1_SEM_multi_cavity} d-i) and estimate the coupling strengths from the separation of the super-modes in the simulated spectra, assuming cavity operation in the range of QD emission ($\sim 900-930$ nm). For a hole radius $r$ varying from $70$ nm down to $50$ nm, with photonic crystal lattice constant $a = 264$ nm, we find that $t/2\pi \sim 0.8-1.3$ THz; $J_1/2\pi \sim 0.4-0.8$ THz and $J_2 << t$,$J_1$.

\begin{figure}
\includegraphics[width=3.5in]{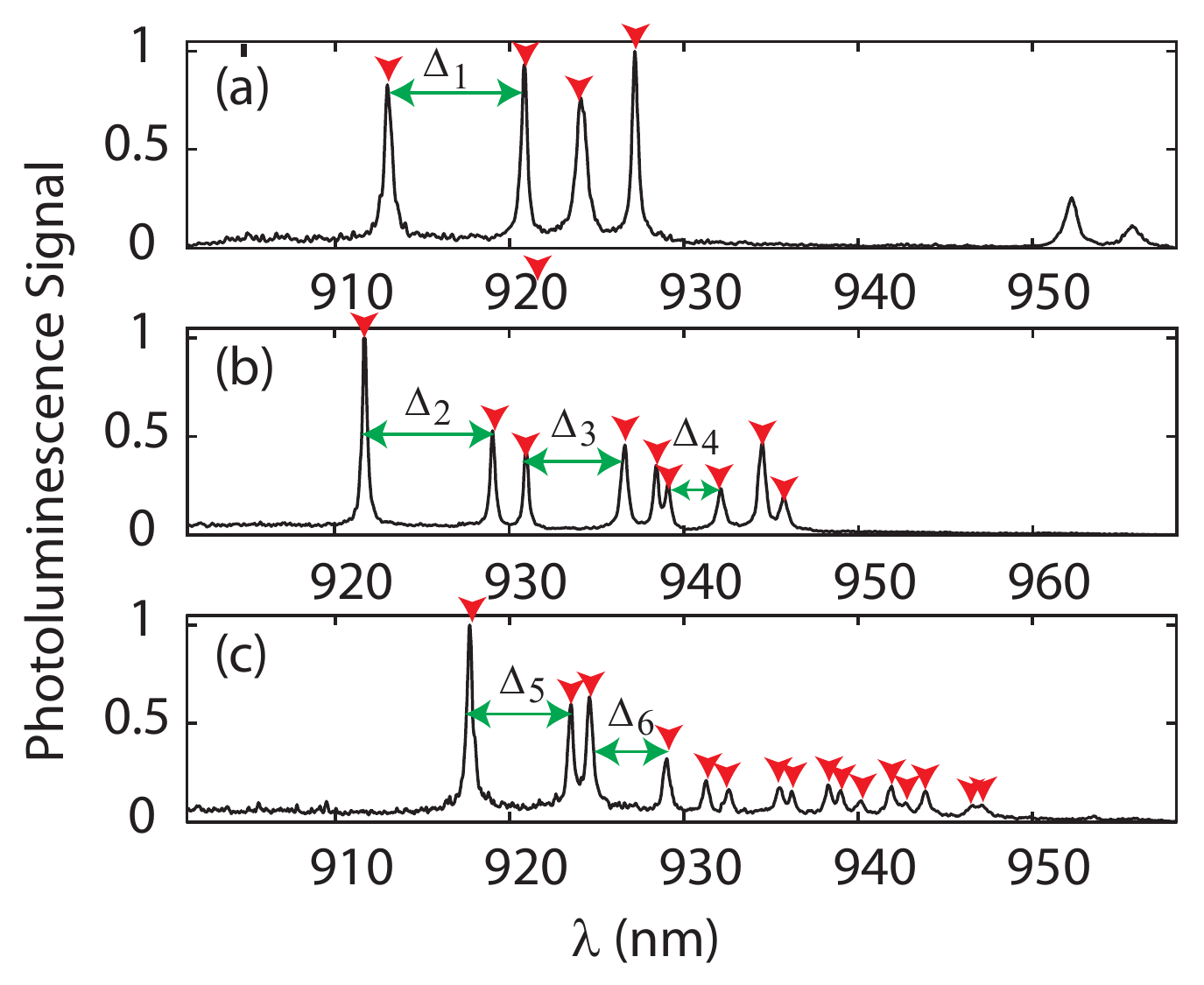}
\caption{(color online) The PL spectra of the CCA for (a) $4$; (b) $9$ and (c) $16$ cavities. We can clearly identify all the cavity array modes. We focus on several specific separations between the CCA modes labeled $\Delta_1$ through $\Delta_6$ in the plots and perform statistical analysis. We also observed several low-Q modes at long wavelengths for several cavity arrays, as can be seen in part (a). These modes are not the actual coupled cavity modes under study, which is confirmed by monitoring the resonance frequencies of single (uncoupled) $L3$ cavities fabricated in the same chip.
\label{Fig3_PL_spectra}}
\end{figure}

We characterize the resonances of the coupled cavity array by photoluminescence (PL) studies, where the large density of embedded QDs acts as an internal light source. Fig. \ref{Fig3_PL_spectra} shows the PL spectra obtained from the CCAs. The spectrum for different CCAs are taken from different e-beam doses, and hence correspond to slightly different hole radii. The quality factors of the observed modes are $\sim 1000-3000$, and all the modes are linearly polarized with similar polarization axis. We note that the set of higher Q-factor resonances in Fig. \ref{Fig3_PL_spectra} are identified as the coupled fundamental modes of the $L3$ cavities, shown in Fig. \ref{Fig1_SEM_multi_cavity}. These modes are not necessarily in the same wavelength range for different sizes of the arrays, as the structures were defined during the fabrication process with different doses in e-beam lithography, and thus photonic crystals have different parameters. We also point out that the number of modes observed in PL should be the same as the number of cavities in the CCA, irrespective of whether the cavities are coupled or not (assuming any degeneracy is lifted due to fabrication imperfection). Without coupling between the cavities, the observed modes would be randomly placed and no specific order between the modes should be observed. On the other hand, in the presence of the coupling between the cavities, the cavity modes are expected to be spaced at a specific order determined by the coupling strengths. However, due to the disorder introduced during the nano-fabrication process, the exact distribution of the cavity resonance frequencies will be perturbed. Hence from a statistical study of the differences in the cavity resonance frequencies we can estimate the ratio between the cavity coupling strengths and the disorder in the cavity resonances. We note that one could instead estimate the disorder in the cavity resonances from the actual cavity frequencies, and not the differences. However, cavities written on different parts of the chip are more susceptible to fabrication variation, and might suffer an overall frequency shift. Thus, the mode separations provide a better measure of the disorder present within each CCA, while allowing us to gather statistics from several CCAs for comparison.

We find a consistent order between the modes of different CCAs (Fig. \ref{Fig3_PL_spectra}), indicating the cavities are coupled. Next we analyze all the separations between the subsequent cavity modes. In order to do this, we fabricated $\sim 30$ copies of each of the three types of cavity arrays, and calculated the mean $\mu$ and standard deviation $\sigma$ of all these mode separations. We note that the mean $\mu$ of the mode separations is a combination of the coupling strength and the disorder, whereas the standard deviation $\sigma$ of the mode separations depends mostly on the disorder, as explained earlier. To elaborate further, we can consider the simple example of a photonic molecule (two coupled cavities), where the observed separation $\Delta$ between two modes is $\sqrt{\Delta_0^2+4J^2}$ with $\Delta_0$ being the random bare detuning between the cavities due to fabrication imperfection and $J$ being the the coupling strength \cite{phot_mol_AM}. We assume that the bare detuning follows a Gaussian probability distribution with zero mean and standard deviation $\sigma_f$, i.e., the probability of having a detuning $\Delta_0$ is
\begin{equation}
\label{eq_gauss}
Pr(\Delta_0)=\frac{1}{\sqrt{2\pi}\sigma_f}e^{-\frac{\Delta_0^2}{2\sigma_f^2}}
\end{equation}
Under such a Gaussian approximation \cite{houck_cqed_paper}, we find that the mean $\mu$ of the mode separation $\Delta$ (we consider the absolute value of the separation) is $\mu=\sqrt{\frac{2}{\pi}}\sigma_f$ if there is no coupling ($J=0$ or (i.e, $\sigma_f/J>>1$) and $\mu=2J+\sigma_f^2/4J+\mathcal{O}(\sigma_f^4)$ if the disorder is weak compared to the coupling i.e., $\sigma_f/J<<1$). The standard deviation $\sigma$ for the mode separations is $\sigma \sim \sigma_f$ without any coupling ($J=0$) and $\sigma \sim \mathcal{O}(\sigma_f^2/J)$ when $\sigma_f/J<<1$. Similar analysis can be performed for CCAs with more than two cavities, although the expressions for the mean and standard deviation become complicated, and a simple closed form expression is difficult to obtain. Nevertheless, as seen for the photonic molecule, the ratio of the standard deviation to the mean gives us an estimate of the relative contribution of the disorder and the coupling to the mode separations. In our fabricated CCAs, we find that the ratio $\sigma/\mu << 1$ for almost all the mode separations, indicating the presence of strong inter-cavity coupling; otherwise, for no coupling, the ratio would be equal to $\sqrt{\pi/2-1} \sim 1$. Table \ref{table_data} shows the data for specific separations ($\Delta_1 \rightarrow \Delta_6$) between the cavity modes in the cavity array spectra in detail. We note that all the separations are not equally influenced by the coupling strengths as seen from the numerical simulations presented below (Fig. \ref{fig_theory}), and the chosen separations (indicated in Fig. \ref{Fig3_PL_spectra}) are the ones that are most heavily influenced by the coupling strengths.

\begin{table}
  \centering
  \caption{The mean mode separations ($\mu$), and standard deviation ($\sigma$) measured over $\sim 30$ cavity arrays, with similar hole radii (see Fig. \ref{Fig3_PL_spectra} for definition of the separations). }\label{table_data}
  \begin{tabular}{c c c c }
  \hline
 $\Delta$ &  $  \hspace{0.2in} \mu (THz)$ &  \hspace{0.2in} $\sigma (THz)$ &  \hspace{0.2in} $\sigma/\mu$ \\
  \hline
 $\Delta_1$ &  \hspace{0.2in} $2.33$& \hspace{0.2in} $0.25$& \hspace{0.2in} $0.1$\\
 $\Delta_2$ &  \hspace{0.2in} $3.22$& \hspace{0.2in} $0.13$& \hspace{0.2in} $0.04$\\
 $\Delta_3$ &  \hspace{0.2in} $2.35$& \hspace{0.2in} $0.14$& \hspace{0.2in} $0.06$\\
 $\Delta_4$ &  \hspace{0.2in} $1.19$& \hspace{0.2in} $0.19$& \hspace{0.2in} $0.16$\\
 $\Delta_5$ &  \hspace{0.2in} $1.94$& \hspace{0.2in} $0.2$& \hspace{0.2in} $0.1$\\
 $\Delta_6$ &  \hspace{0.2in} $2.35$& \hspace{0.2in} $0.14$& \hspace{0.2in} $0.06$\\
  \hline
\end{tabular}
\end{table}

\section{Estimation of coupling and disorder}
\begin{figure}
\includegraphics[width=3.5in]{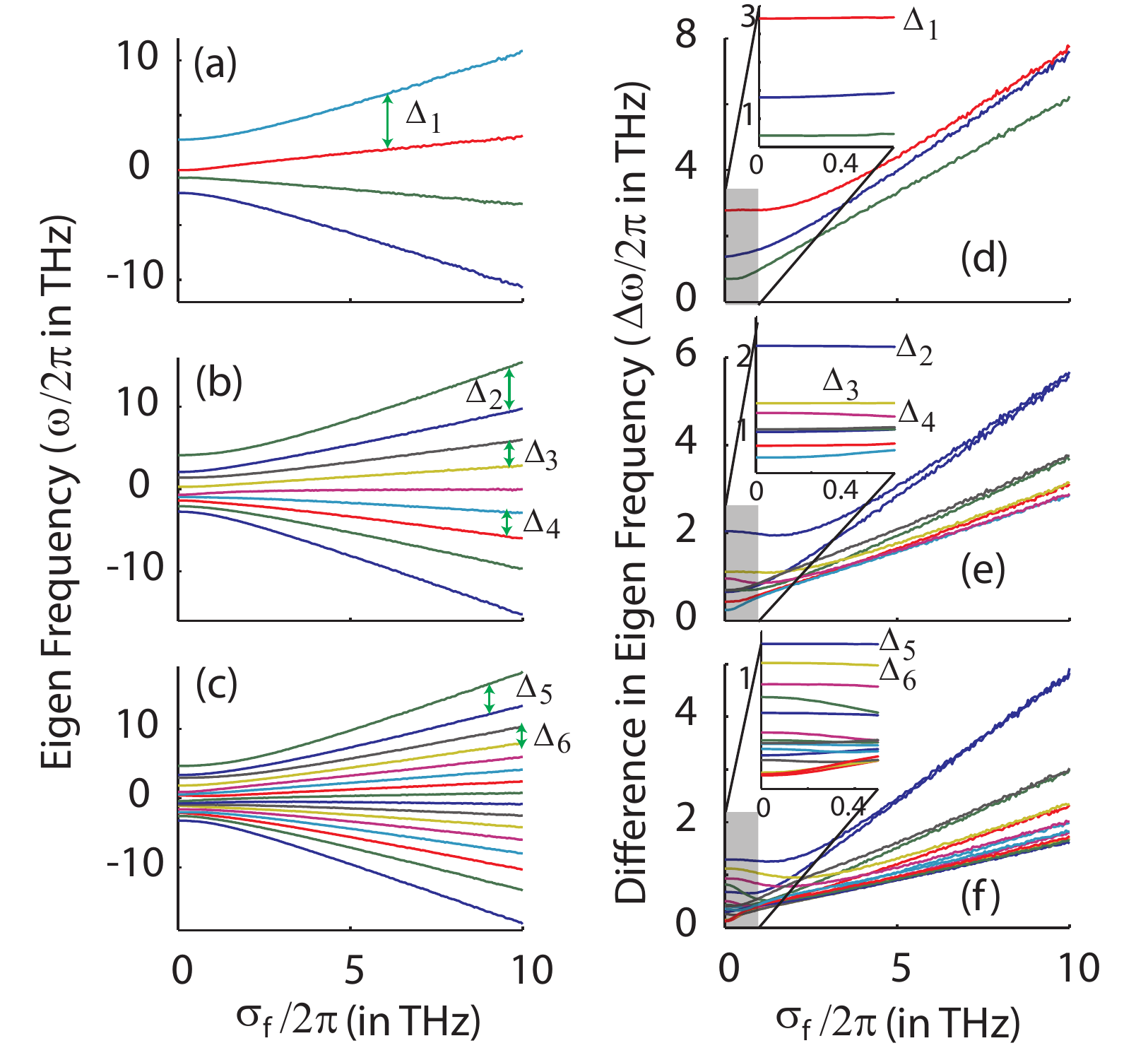}
\caption{(color online) Numerically-calculated eigen-spectra of the coupled cavities: the eigen-frequencies as a function of the disorder standard deviation $\sigma_f$ for (a) $4$, (b) $9$ and (c) $16$ cavities in the arrays. The spacings between the cavities are the same as the structures shown in the SEM images in Figs. \ref{Fig1_SEM_multi_cavity} a-c. The difference in the subsequent eigen-values are shown as a function of $\sigma_f$ for (d) $4$, (e) $9$ and (f) $16$ cavities in the arrays. We note that the mode separations increase linearly with increasing $\sigma_f$, when $\sigma_f$ is much greater than the coupling strengths, as found in the theory from a simple photonic molecule. Insets magnify the region of low disorder and we identify the mode separations $\Delta_1 \rightarrow \Delta_ 6$.
\label{fig_theory}}
\end{figure}

Using the coupling strengths derived from FDTD simulations ($t/2\pi =1.2$ THz, $J_1/2\pi=0.8$ THz, $J_2 \approx 0$), we calculate the eigen-states of the CCA by diagonalizing the Hamiltonian $\mathcal{H}$:
\begin{equation}
\mathcal{H}=\sum _i \Delta_i a_i^\dag a_i+\sum_{\langle i,j \rangle} g_{ij} (a_i^\dag a_j+a_j^\dag a_i)
\end{equation}
where, $\Delta_i$ is the resonance frequency of the $i^{th}$ cavity due to fabrication imperfection, and $g_{i,j}$ is the coupling strength between the $i^{th}$ and $j^{th}$ cavity. The cavity frequencies $\Delta_i$'s are randomly chosen from a Gaussian distribution with zero mean and standard deviation $\sigma_f$ (given by Eq. \ref{eq_gauss}). We assume that the coupling strengths are constant and do not depend on the disorders. The mean values of the eigen-frequencies (averaged over $\sim 10000$ instances) are plotted in Figs. \ref{fig_theory}a,b,c as a function of increasing $\sigma_f$. We observe that the relative detunings between the modes follow a specific order when the disorder is small. However, with increasing disorder any specific order between the cavity modes disappears. This can be observed more clearly in Figs. \ref{fig_theory}d,e,f, where the differences in the mode frequencies are plotted as a function of $\sigma_f$. We note that the differences become similar, and increase linearly with $\sigma_f$, as found in the simple photonic molecule model previously. We also note that several modes are spaced very closely at weak disorder, indicating a lesser contribution of the coupling to such mode separations (inset of Figs. \ref{fig_theory}d,e,f). On the other hand several detunings between the modes are large compared to others (denoted by $\Delta_1 \rightarrow \Delta_ 6$) signifying a large contribution from the coupling strengths to the mode separations. We observe that the relative positions of the cavity modes match qualitatively with our experimental findings.  We can identify the specific separations $\Delta_1 \rightarrow \Delta_ 6$ between the modes (Fig. \ref{fig_theory}). Clearly, the fabricated structures are in the regime where the coupling strengths are greater than the disorder. This regime is magnified in the inset of Figs. \ref{fig_theory}d,e,f; and the mode separations $\Delta_1 \rightarrow \Delta_ 6$ are identified.

\begin{figure}
\includegraphics[width=3.5in]{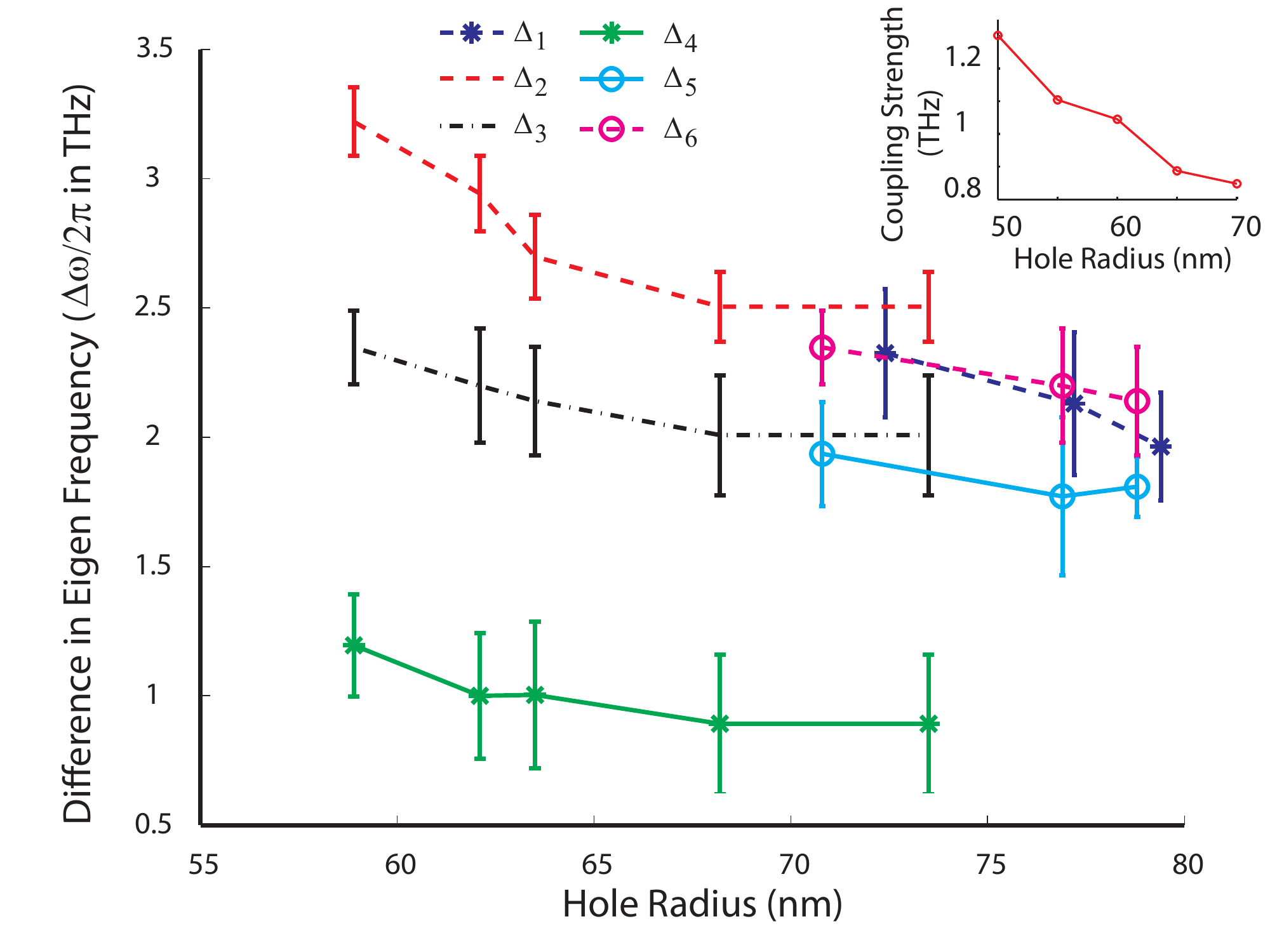}
\caption{(color online) The mode separations $\Delta_1 \rightarrow \Delta_6$ as a function of the photonic crystal hole radius. A decreasing trend in the separation is observed with the increasing hole radius (the photonic crystal lattice periodicity $a$ is $264$ nm). For comparison, the inset shows the numerically simulated (FDTD) coupling strength between two cavities placed diagonally as a function of the hole radius.
\label{fig4_coupling}}
\end{figure}

Finally, as a further proof of the fact that the detunings between the observed cavity array modes are mostly due to the coupling between the cavities, and not the disorder, we repeated the fabrication of sets of $\sim 30$ cavities for different values of air-hole radius for all three types of CCAs. A decreasing trend in the separation is observed with increasing hole radius (Fig. \ref{fig4_coupling}). A similar trend is observed in simulation for a photonic molecule with diagonally placed cavities, as a function of the hole radius (inset of Fig. \ref{fig4_coupling}). Such a trend also indicates that the separations are mostly due to the coupling between cavities, as a detuning due solely to disorder would have a much weaker dependence on the photonic crystal hole size. The decrease in the mode separation with increase in the hole radius can be explained by the increase in the photonic band gap size with increasing hole radius (and thus larger reflectivity of the mirror layers separating cavities, which reduces the cavity couplings).

\section{Conclusion}
We show the signature of large coupling strengths between photonic crystal cavities, in a coupled cavity array fabricated in GaAs containing InAs QDs. We observe that the coupling strengths are significantly larger than the disorder introduced during the nano-fabrication. Satisfying this condition is necessary for employing such cavity arrays in quantum simulation with correlated photons, although the challenge of achieving a nonlinearity in each cavity still remains open.

The authors acknowledge financial support provided by the Office
of Naval Research (PECASE Award; No: N00014-08-1-0561), DARPA
(Award No: N66001-12-1-4011), NSF (DMR-0757112) and Army Research
Office (W911NF-08-1-0399). Work at UT-Austin was supported by the Air Force Office of Scientific Research $(YIP: FA9550-10-1-0182)$. A.R. is also supported by a Stanford Graduate Fellowship.This work was performed in part at the Stanford Nanofabrication and University of Texas Microelectronics Research Center facilities of the NNIN, supported by the National Science Foundation.

\bibliography{CCA}
\end{document}